\documentclass[a4paper,11pt]{article}
\usepackage{pos}
\usepackage{caption} 
\usepackage{subcaption} 
\usepackage{tikz}
\usepackage{braket}
\usepackage{pgfplots} 
\pgfplotsset{compat=1.17} 
\newcommand{\ie}{{\it i.e.} }

\title{de Sitter Connectivity from Holographic Entanglement}

\author*[a]{Victor Franken}

\affiliation[a]{CPHT, CNRS, Ecole polytechnique, IP Paris,\\
 F-91128 Palaiseau, France}

\emailAdd{victor.franken@polytechnique.edu}

\abstract{
Considering two antipodal observers in de Sitter space, we illustrate how spacetime connectivity between the holographic screens located on the (stretched) horizons emerges from holographic entanglement. To do so, we construct a covariant holographic entanglement entropy prescription in de Sitter space, including quantum corrections. While this prescription is inspired by the bilayer proposal, we argue that the monolayer proposal appears to be inconsistent. Entanglement wedge reconstruction implies an extension of static patch holography where the exterior region connecting the static patches is included, and reconstructible from the two screens. A phase transition occurs where there is an exchange of dominance between two competing quantum extremal surfaces, leading to a transfer of the encoding of the exterior region from one screen to the other. The effects in the bulk of integrating out degrees of freedom from the screens are discussed.
}

\FullConference{Corfu Summer Institute 2023 "School and Workshops on Elementary Particle Physics and Gravity" (CORFU2023)\\
 23 April - 6 May , and 27 August - 1 October, 2023\\
Corfu, Greece\\}

\begin{document}
\maketitle

\section{Introduction}
\label{sect:intro}

\begin{figure}[h!]
\begin{subfigure}[t]{0.48\linewidth}
\centering
\begin{tikzpicture}[scale=0.60]
			\begin{scope}[transparency group]
		\begin{scope}[blend mode=multiply]
\path
       +(3.5,3) coordinate (IItopright)
       +(-3.5,3) coordinate (IItopleft)
       +(3.5,-3) coordinate (IIbotright)
       +(-3.5,-3) coordinate(IIbotleft)
       +(-0.5,0) coordinate (IIcenterleft)
       +(+0.5,0) coordinate (IIcenterright)
      
       ;
       
        \fill[fill=blue!20] (-3.5,3) -- (-3.5,-3) -- (-0.5,0) -- cycle;
        \fill[fill=blue!20] (3.5,3) -- (3.5,-3) -- (0.5,0) -- cycle;

\draw[thick] (IItopright) --
          node[midway, above, sloped, thick] {}
      (IIbotright);

\draw[thick] (IIbotleft) --
          node[midway, above , sloped, thick] {}
      (IItopleft);
      
\node at (0,0) [label = center:{$\otimes$}]{};

\draw (IItopleft) -- (IIcenterleft) -- (IIbotleft) ;
\draw (IItopright) -- (IIcenterright) -- (IIbotright) ;
\end{scope}
\end{scope}
\end{tikzpicture}
\caption{}
\end{subfigure}
\begin{subfigure}[t]{0.48\linewidth}
\centering
\begin{tikzpicture}[scale=0.60]
			\begin{scope}[transparency group]
		\begin{scope}[blend mode=multiply]
\path
       +(3,3) coordinate (IItopright)
       +(-3,3) coordinate (IItopleft)
       +(3,-3) coordinate (IIbotright)
       +(-3,-3) coordinate(IIbotleft)
      
       ;
       
       \fill[fill=red!20] (IItopleft) decorate[decoration=snake] {[bend right=20] to (IItopright)} -- (0,0);
       
       \fill[fill=red!20] (IIbotright) decorate[decoration=snake] {[bend right=20] to(IIbotleft)} -- (0,0);

       \fill[fill=blue!20] (IItopleft) -- (IIbotleft) -- (0,0);

       \fill[fill=blue!20] (IItopright) -- (IIbotright) -- (0,0);
       
\draw[decorate,decoration=snake] (IItopleft) 
          node[midway, above, sloped]    {} [bend right=20] to
      (IItopright);
      
\draw[thick] (IItopright) --
          node[midway, above, sloped] {}
      (IIbotright);
      
\draw[decorate,decoration=snake]  (IIbotright)
          node[midway, below, sloped] {} [bend right=20] to
      (IIbotleft);
      
\draw[thick] (IIbotleft) --
          node[midway, above , sloped] {}
      (IItopleft);

\draw (IItopleft) -- (IIbotright)
        (IItopright) -- (IIbotleft) ;
    \end{scope}
\end{scope}
\end{tikzpicture}
\caption{}
\end{subfigure}
    \caption{Gravity dual to the thermofield-double state. (a) At low temperature, the TFD is dual to two disconnected AdS spactimes. (b) At high temperature, the bulk dual is a maximally extended AdS-Schwarzschild black hole, with a wormhole connecting the two CFTs. The two initial spacetimes are depicted in blue and the Einstein-Rosen bridge is depicted in red.}
    \label{fig:ER=EPR}
\end{figure}
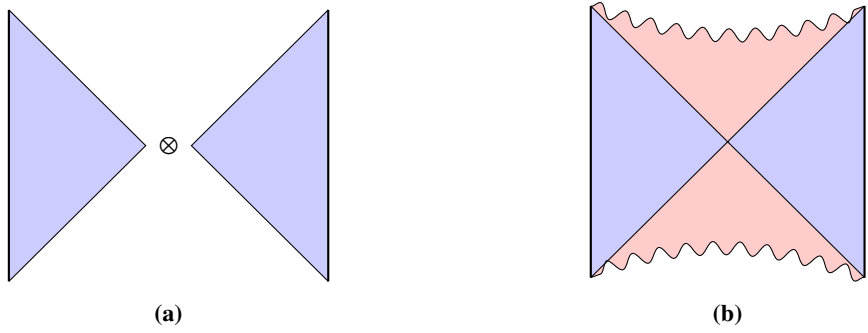

The AdS/CFT correspondence has led to great progresses in the understanding of black hole information and the connection between spacetime and entanglement (see e.g. \cite{Maldacena:2001kr,VanRaamsdonk:2009ar,VanRaamsdonk:2010pw, Maldacena:2013xja,Penington:2019npb,Almheiri:2019psf,Almheiri:2019hni,Engelhardt:2023xer}). A typical example of this relation is the emergence of spacetime from the entanglement between two disconnected holographic CFTs in a thermofield-double (TFD) state. The TFD is a pure state and tracing out one of the two sides produces the original CFT in a thermal state. At small temperatures, corresponding to low entanglement between the two CFTs, the gravity dual to the TFD state corresponds to two disconnected spacetimes. On the other hand, at high temperatures where the two CFTs are highly entangled, the dual spacetime consists of an eternal AdS black hole. As pictured in figure \ref{fig:ER=EPR}, the two initial spacetimes are connected by an Einstein-Rosen (ER) bridge in the latter case. The idea that spacetime connectivity between two holographic theories emerges from entanglement is often associated to the slogan ``ER=EPR'' \cite{Maldacena:2013xja}.\footnote{For a proposition of formal definition of this equivalence, see \cite{Engelhardt:2023xer}.}
In parallel, a precise holographic dictionary between entanglement in the non-gravitating theory and bulk quantities was developed by Ryu, Takayanagi, and others \cite{Ryu:2006bv,Hubeny:2007xt,Faulkner:2013ana, Engelhardt:2014gca,Wall:2012uf}. This led to the development of subregion-subregion duality, or entanglement wedge reconstruction, which allows to determine the bulk subregion dual to a given subregion of the quantum theory \cite{Dong:2016eik, Cotler:2017erl,
Leutheusser:2022bgi}.

In this paper we suppose that the holographic nature of gravity and the notions of holographic entanglement entropy and entanglement wedge reconstruction should be universal \cite{tHooft:1993dmi,Susskind:1994vu, Bousso:2002ju}, at least in the semiclassical limit $G_N\rightarrow 0$, and thus applicable to cosmological models, in particular to the de Sitter spacetime. We interpret the area of extremal surfaces in the bulk as first order contributions to the entropy of subsystems in the non-gravitating dual quantum theory, and bulk fields semiclassical entropy as second order contributions, see Section \ref{sec:ent}. These extremal surfaces should satisfy the usual homology constraint \cite{Hubeny:2007xt} and their associated entropy and entanglement wedge should satisfy weak and strong subadditivity \cite{Wall:2012uf} as well as entanglement wedge nesting \cite{Akers:2016ugt}.

 Although the task of finding an explicit holographic duality in de Sitter space turns out to be complicated, partially for the lack of a string theory description of de Sitter space, it has attracted a lot of attention lately \cite{Anninos:2011ui, Hikida:2022ltr, Banihashemi:2022htw, Susskind:2021omt, Alishahiha:2004md, Alishahiha:2005dj, Dong:2010pm, Dong:2018cuv, Arias:2019pzy, Arias:2019zug, Emparan:2022ijy, Svesko:2022txo, Hartman:2020khs, Balasubramanian:2020coy, Balasubramanian:2020xqf, Geng:2021wcq, Sybesma:2020fxg, Aalsma:2021bit,Kames-King:2021etp, Levine:2022wos, Narayan:2015vda, Narayan:2017xca, Chapman:2021eyy,Shaghoulian:2021cef, Shaghoulian:2022fop, Susskind:2021esx, Susskind:2022dfz, Susskind:2022bia, Rahman:2022jsf, Goel:2023svz, Narovlansky:2023lfz, Rahman:2023pgt, Susskind:2023rxm, Balasubramanian:2023xyd, Giveon:2023rsk, Franken:2023pni, Kawamoto:2023nki, Galante:2023uyf}. In \cite{Franken:2023pni}, based on a recent holographic conjecture in de Sitter space \cite{Susskind:2021omt}, we considered the possibility to define a covariant holographic entropy prescription in de Sitter space as well as its consequence concerning entanglement wedge reconstruction. 
 

The presence of holographic degrees of freedom on the cosmological horizon (see Figure \ref{fig:dS}) leads to a number of constraints concerning the construction of a holographic entanglement entropy formula. Two approaches to this problem were considered by Susskind and Shaghoulian \cite{Susskind:2021esx, Shaghoulian:2021cef, Shaghoulian:2022fop}, named the monolayer and bilayer proposal. We argue that generic assumptions make the monolayer proposal inconsistent and propose a covariant prescription inspired by the bilayer proposal. This exposes intricacies previously overlooked in the monolayer and bilayer proposals of Susskind and Shaghoulian, resulting in significant differences when quantum corrections are taken into account. Moreover, it allows for an extension of the prescription to a general class of closed FRW cosmologies \cite{Franken:2023jas}. Using this prescription we are able to study entanglement wedge reconstruction which leads to the two main takeaways of this work. First, the only consistent definition of extremal surfaces in de Sitter space implies that the connectiveness of de Sitter space emerges from the entanglement between the causal patches of two antipodal observers. In particular, the spatial region connecting these two patches can be reconstructed from the holographic degrees of freedom. The second one is that there are two competing quantum extremal surfaces when computing the entropy of one of the two screens, associated to two drastically different entanglement wedges. When the first one dominates, the entanglement wedge of one screen only covers its interior region, corresponding to the usual static patch holography picture. When the second dominates, the entanglement wedge also extends in the exterior and connects the two screens via a wormhole. The dominant quantum extremal surface can undergo a phase transition from one to the other, corresponding to a transfer of the encoding of the wormhole connecting the two holographic screens. To conclude, we discuss the implications in the bulk of integrating out degrees of freedom from the screens, and how the encoding of the exterior region is lost when the screens are pushed inside the static patches at a distance greater than the Planck scale from the cosmological horizons. Further details on the following discussions as well as additional results may be found in \cite{Franken:2023pni, Franken:2023jas}.
\section{A Holographic Description of de Sitter Space?}
In this section we will motivate the holographic picture on which we will base the rest of our paper. Roughly, we shall motivate that any physically meaningful description of the universe is intrinsically observer dependent, naturally leading to study the region causally connected to some observer worldline. In an inflationary cosmology - de Sitter space - this region is bounded by a cosmological horizon. We will then show how, analogously with the AdS/CFT correspondence, holographic degrees of freedom located on this boundary are expected to encode the complete state of the causal region of the observer. The connectivity of the full universe can then be related to the entanglement between the holographic degrees of freedom associated to causally disconnected observers.
\subsection{The Observer and the Static Patch}
\label{sec:observer}
Since the 90's, there have been strong indications that a quantum theory of gravity should present holographic properties, meaning that the theory is described by degrees of freedom located on the boundary of space \cite{tHooft:1993dmi, Susskind:1994vu, Bousso:2002ju}. This conjecture was explicitly realized by the AdS/CFT correspondence in the context of string theory \cite{Maldacena:1997re, Witten:1998qj}. One would naturally be interested to test this correspondence in our own universe. However, both the inflationary period and the late time universe are in an accelerated expansion phase, modeled by the de Sitter space which has a closed topology, \ie no boundaries. There are three ways to interpret the absence of spatial boundary. The first one is to conclude that the de Sitter space does not have any holographic description, which would be highly unsatisfactory and inconsistent with the Bousso covariant entropy conjecture \cite{Bousso:1999xy}. The second is to seek for a modification of the original formulation of holography, as Strominger did by proposing the dS/CFT correspondence \cite{Strominger:2001gp} with holographic degrees of freedom located at the future infinity, timelike conformal boundary. The third one, which may be the simplest one, would be to simply state that the de Sitter space does not have any dynamical degree of freedom \cite{Susskindconf, Shaghoulian:2023odo}. Here we consider the latter as it is well motivated by gravitational path integrals and swampland conditions \cite{Shaghoulian:2023odo, McNamara:2020uza, Marolf:2020xie, Almheiri:2019hni}. For example, the island formula \cite{Almheiri:2019hni, Penington:2019kki, Almheiri:2019qdq} implies that there is no way to entangle a closed universe with any complementary system \cite{Shaghoulian:2023odo}, leading to the conclusion that state of the universe is unique.

Another argument which suggests that the Hilbert space of de Sitter space is trivial is that the algebra of observables in de Sitter space is trivial \cite{Chandrasekaran:2022cip, Witten:2023xze}. The only possibility to get a sensible algebra is to include the observer explicitly in the system, which effectively breaks the de Sitter symmetries as it selects one particular static patch. This produces an algebra of type II$_1$ which has a state of maximal entropy corresponding to an empty static patch tensored with a simple model of observer (essentially a clock). The non-triviality of the algebra in the presence of an observer is analogous to the potentially very complex state that may be measured by an observer, despite the triviality of the state of the full universe.  This is somewhat reminiscent of topological field theories, like Chern-Simons theory on a closed manifold, whose Hilbert space is trivial.\footnote{We thank Lior Benizri and Jan Troost for pointing this out to us.} In such cases, degrees of freedom are effectively added when considering subsystems, which has the effect of relaxing some of the gauge invariance (the ones broken by the subsystem boundary).

Let us work under the assumption that the Hilbert space of de Sitter space has a unique pure state. Since a closed universe cannot be observed from the outside, an explicit and non-trivial description of de Sitter space should include some observer. This observer can measure fields along its worldline, and generally may interact with anything inside its causal patch. The discussion above motivates the fact that it is natural to consider the static patch of an observer in order to probe non-trivial behaviors of gravity in de Sitter space. In particular, the static patch is self-contained in the sense that an observer may neglect everything happening outside the cosmological horizon. The region of the universe he observes is effectively bounded by its cosmological horizon. So that, from his point of view, the observable universe has a boundary on which it would seem sensible to place holographic degrees of freedom. Moreover, it has been argued that the horizon of the observer, which is the boundary of the static patch, is associated to an entropy proportional to its area \cite{Gibbons:1977mu},
\begin{equation}
	S_{\rm GH}=\frac{A}{4G\hbar}.
\end{equation}
The fact that thermodynamic quantities can be associated to the horizon of the static patch is indicative of an holographic behavior. In particular one could expect that degrees of freedom located on the cosmological horizon should be able to encode everything happening in some region of de Sitter space, but which one? The discussion above indicates that an intuitive answer would be the static patch, where the degrees of freedom describing it emerge from the choice of worldline associated to the static patch observer. In the next section, we formally motivate this using the Bousso bound \cite{Bousso:1999xy}.

\subsection{The Holographic Principle and de Sitter Space}
\label{sec:dSholo}

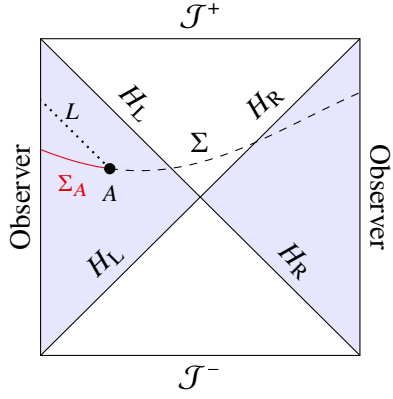
\begin{figure}[h!]
	\centering
	\begin{tikzpicture}[scale=0.7]
		\begin{scope}[transparency group]
			\begin{scope}[blend mode=multiply]
				\path
				+(3,3)  coordinate (IItopright)
				+(-3,3) coordinate (IItopleft)
				+(3,-3) coordinate (IIbotright)
				+(-3,-3) coordinate(IIbotleft)
				
				;
				\draw(IItopleft) --
				node[midway, above, sloped] {$\mathcal{J}^+$} (IItopright) --
				node[midway, above, sloped] {Observer} (IIbotright)--
				node[midway, below, sloped] {$\mathcal{J}^-$} (IIbotleft) --
				node[midway, above, sloped] {Observer}(IItopleft) -- cycle;
				
				\fill[fill=blue!10] (-3,3) -- (-3,-3) -- (0,0);
				
				\fill[fill=blue!10] (3,3) -- (3,-3) -- (0,0);

				\draw[dotted, thick] (-1.7,0.538) -- (-3,1.838);
				
				\draw[domain=-3:-1.7, smooth, variable=\x, red] plot ({\x}, {sin(deg((\x/2-1)))+1.5});
				\draw[domain=-1.7:3, smooth, variable=\x, black, dashed] plot ({\x}, {sin(deg((\x/2-1)))+1.5});
				
				\node at (-2.4,0.8) [label = below:\footnotesize{\color{red}$\Sigma_A$}]{};
				\node at (-1.7,0.538) [circle, fill, inner sep=1.5 pt, label = below:\footnotesize{$A$}]{};\node at (-2.4,1.1) [label = above:\footnotesize{$L$}]{};

				\draw  (IItopleft) --  node[midway, above, sloped] {$H_{\rm L}$} (0,0)-- node[midway, above, sloped] {$H_{\rm R}$} (IIbotright)
				(IItopright) --node[midway, above, sloped] {$H_{\rm R}$} (0,0) -- node[midway, above, sloped] {$H_{\rm L}$} (IIbotleft) ;
				
				\node at (0,1.6) [label=below:$\Sigma$]{};
				
			\end{scope}
		\end{scope}
	\end{tikzpicture}
\caption{Two antipodal observers follow the antipodal trajectories defined by the left and right edge of the diagram. The diagonal lines form the cosmological horizons $H_{\rm L}$ and $H_{\rm R}$ for the two observers. They bound the two static patches $P_{\rm L}$ and $P_{\rm R}$ depicted in blue. The dashed line is an arbitrary global spacelike slice one which we place a sphere $A$ depicted by a black dot. $\Sigma_A\in\Sigma$ is the red segment and the future lightsheet emanating from $A$ is the dotted line. \label{fig:dS}.}
\end{figure}

The geometry of de Sitter space and an example of global spacelike  slice is depicted in Figure \ref{fig:dS}. Every complete spacelike slice of $(n+1)$-dimensional de Sitter space has the topology of a $n$-sphere. Under time evolution, this $n$-sphere  undergoes a phase of contraction followed by a phase of expansion, reaching infinite size at time infinities $\mathcal{J}^{\pm}$.
 Let us consider an observer whose worldline follows a timelike geodesic in the bulk. All geodesics are equivalent under the $SO(1,n+1)$ symmetry group of de Sitter space, so we can choose a coordinate system in which the observer worldline follows the left edge of the diagram. This corresponds to a static observer at the north pole (or pode) of the $n$-sphere. The region of spacetime that can interact with the observer is called the static patch $P_{\rm L}$ and is delimited by a past and a future horizon, forming $H_{\rm L}$. There is a unique worldline such that the static patch $P_{\rm R}$ of an observer following this trajectory does not overlap with $P_{\rm L}$, corresponding to the right edge of the Penrose diagram, describing a static observer on the south pole, or antipode.
 
The static patch holography conjecture is motivated by the covariant entropy bound and the holographic principle of Bousso \cite{Bousso:1999xy,Bousso:2002ju}. To understand this, we start with an arbitrary Cauchy slice $\Sigma$. Let $A$ be a closed codimension $2$ surface inside $P_{\rm L}$. The following discussion also applies to $P_{\rm R}$  with associated horizon $H_{\rm R}$ by symmetry. The surface $A$ defines the boundary of two subsystems of $\Sigma$: one which is entirely contained in $P_{\rm L}$, denoted $\Sigma_A$, and the complement of $\Sigma_A$ with respect to $\Sigma$ which intersects $H_{\rm L}$ and $H_{\rm R}$, see Figure \ref{fig:dS}. The Cauchy slice $\Sigma$ can evolve under unitary evolution. In  particular, one can consider the unitary transformations of $\Sigma$ such that $A\in\Sigma$ is a fixed point. This defines the evolution of $\Sigma_A$ with its boundary $A$ fixed. In particular, we can consider the limit of future-directed unitary evolution of $\Sigma_A$ which goes to a future-directed lightlike surface emanating from $A$ towards the pode, noted $L$. $L$ may also be constructed by considering the congruence of future-directed lightrays emanating from $A$ and going towards the pode. The second law of thermodynamics implies that
\begin{equation}
	\label{eq:ineqL}
	S_{\text{coarse}}(\Sigma_A)\leq S_{\text{coarse}}(L),
\end{equation}
where $S_{\text{coarse}}$ describes the coarse-grained, or thermodynamic entropy which corresponds to the number of degrees of freedom necessary to the description of the quantum state of a system. The null congruence $L$ is entirely contained in $P_{\rm L}$ by definition. Any null congruence contained in $P_{\rm L}$ and directed towards the pode in $P_{\rm L}$ is a lightsheet, \textit{i.e.} is of non-positive expansion (see for example \cite{Franken:2023pni}). Therefore, the covariant entropy bounds \cite{Bousso:1999xy} applies and
\begin{equation}
	\label{eq:Bousso}
	S_{\text{coarse}}(L)\leq \frac{\text{Area}(A)}{4G\hbar}.
\end{equation}
Combining equations \eqref{eq:ineqL} and \eqref{eq:Bousso} gives the inequality
\begin{equation}
	\label{eq:B}
	S_{\text{coarse}}(\Sigma_A)\leq  \frac{\text{Area}(A)}{4G\hbar}.
\end{equation}
Hence, a quantum theory of gravity in de Sitter space should account for the property that the number of degrees of freedom needed to describe a spatial region $\Sigma_A$ of the static patch is bounded from below by the area of $A$ divided by $4G\hbar$. This bound makes very clear that any candidate of underlying theory consistent with the covariant entropy bound should unify the geometrical notions of gravity with the description of matter. The holographic principle allows us to make the bound manifest and states that the region $\Sigma_A$ can be described by a set of degrees of freedom localized on $A$ with less than one degree of freedom per Planck area.

The placement of $A$ as an holographic screen is arbitrary. However, the area of closed surfaces on $\Sigma$ increase as we get closer to the cosmological horizon,\footnote{Past the cosmological horizon, future directed lightrays emitted towards the pode are no longer of non-positive expansion and equation \eqref{eq:B} no longer holds.} such that pushing $A$ towards $H_{\rm L}$ maximizes the number of degrees of freedom that are holographically encoded. Hence, on any spacelike slice $\Sigma$, the surface with the greatest number of holographic degrees of freedom is located at the cosmological horizon of the static patch, and describes the part of $\Sigma$ in the static patch. We call this set of degrees of freedom $\mathcal{S}_{\rm L}$. The previous argument applies to any spacelike slice of the static patch, which leads to the static patch holography conjecture:

 \emph{The quantum theory of gravity in the static patch of an observer in de Sitter space is completely described by a quantum system defined on the cosmological horizon of the observer \cite{Susskind:2021omt}.}\footnote{This can be regulated by placing the screen on a stretched horizon close to the cosmological horizon, see Section \ref{sec:stretched}.}

A precise realization of this duality is still lacking although promising results have been found in lower-dimensional de Sitter space \cite{Shaghoulian:2021cef, Shaghoulian:2022fop, Susskind:2021esx, Susskind:2022dfz, Susskind:2022bia, Rahman:2022jsf, Goel:2023svz, Narovlansky:2023lfz, Rahman:2023pgt}. We will not consider these issues here.

\subsection{Bridging the Static Patches}
\label{sec:bridge}

Let us note $\mathcal{H}_{\rm L}$ the Hilbert space of the quantum system located on the cosmological horizon $H_{\rm L}$ and describing $P_{\rm L}$. The area of the de Sitter cosmological horizon is constant in time, in other words any spacelike slice of $H_{\rm L}$ has the same area. The number of degrees of freedom needed to describe the static patch is therefore constant in time, and $\mathcal{H}_{\rm L}$ has a fixed dimension. We thus expect time evolution of this quantum system to be unitary. As mentioned earlier, one can consider the holographic dual to $P_{\rm R}$, with Hilbert space $\mathcal{H}_{\rm R}$. The union of the two holographic theories describes the two disconnected static patches, $P_{\rm L}\cup P_{\rm R}$ depicted in blue in Figure \ref{fig:dS}. Here we would like to consider the effect of entangling the degrees of freedom in $\mathcal{H}_{\rm L}$ and $\mathcal{H}_{\rm R}$. 

This is reminiscent to the thermofield-double state in AdS, where two non-interacting and disentangled holographic quantum theories are dual to two disconnected bulk spacetimes. Once entangled, they form a state which is dual to a spacetime in which the two original spacetimes are connected by a wormhole. We have a very similar situation here with two quantum theories of Hilbert spaces $\mathcal{H}_{\rm L}$ and $\mathcal{H}_{\rm R}$ independently encoding $P_{\rm L}$ and $P_{\rm R}$, respectively. As in the AdS case, one may wonder what the holographic dual to an arbitrary theory of Hilbert space $\mathcal{H}_{\rm L}\otimes \mathcal{H}_{\rm R}$ is. Note that this region should at least include the two static patches, encoded independently in $\mathcal{H}_{\rm L}$ and $\mathcal{H}_{\rm R}$. When degrees of freedom on both sides are entangled we expect the two bulk duals $P_{\rm L}$ and $P_{\rm R}$ to be spatially connected by a spatial bridge. This is the case in pure de Sitter space which is connected by a contracting or expanding region connecting the two static patches (the white regions in Figure \ref{fig:dS}). In particular, the static patch of an observer is in a thermal state  \cite{Gibbons:1977mu} (see section \ref{sec:2screens} for a reminder of this construction) such that its coarse-grained entropy, given by the Gibbons-Hawking law $A/4G\hbar$, is equal to its entanglement entropy. Hence the states of $\mathcal{H}_{\rm L}$ and $\mathcal{H}_{\rm R}$ should be entangled.

We will confirm this intuition by studying entanglement wedge reconstruction in de Sitter spacetime to show that the region between the two cosmological horizons must be included in the holographic duality and that the entanglement wedge of the holographic screens do cover this exterior region. A first step in this direction is the definition of a covariant prescription to compute entropies is the quantum dual theory. This task unravels subtleties that where not present in the monolayer and bilayer proposals of Susskind and Shaghoulian \cite{Susskind:2021esx, Shaghoulian:2021cef, Shaghoulian:2022fop}, and leads to non-trivial results when including quantum corrections.

\section{Covariant Holographic Entropy in de Sitter Space}
\label{sec:ent}

As we have yet no candidate of holographic dual of the static patch in arbitrary dimension, it would be interesting to investigate the possibility to study holographic entanglement entropy in this picture. The computations are done exclusively in the bulk and may provide us with hints concerning the nature of the holographic dual.\footnote{Similarly, one can study holographic complexity with the same motivations, see e.g. \cite{Reynolds:2017lwq,Chapman:2021eyy, Susskind:2021esx,Susskind:2022dfz,Jorstad:2022mls,Anegawa:2023wrk,Anegawa:2023dad,Baiguera:2023tpt,Aguilar-Gutierrez:2023zqm,Auzzi:2023qbm,Aguilar-Gutierrez:2024rka}} Shaghoulian and Susskind recently considered two prescriptions that they call the monolayer and bilayer proposals. However, we still lacked of a precise covariant procedure, which leads to a number of nontrivial subtleties being missed.

 
We want to find  a bulk prescription capable of computing entanglement entropies of spatial subsystems of $H_{\rm L}\cup H_{\rm R}$, corresponding to a subsystem $\mathcal{A}$ of the holographic screens $\mathcal{S}_{\rm L}\cup\mathcal{S}_{\rm R}$. Here we differentiated between the spatial subregion $A$ with the subsystem  $\mathcal{A}$ consisting of degrees of freedom located on $A$. However, as holographic degrees of freedom are located on $H_{\rm L}\cup H_{\rm R}$ with a specific density of degrees of freedom per unit area, we consider that specifying one defines uniquely the other. The monolayer and bilayer proposals are motivated by the bit thread formulation of holographic entanglement entropy \cite{Freedman:2016zud, Headrick:2022nbe} and by wedge holography \cite{Geng:2020fxl, Mollabashi:2014qfa, Miao:2020oey, Akal:2020wfl, Bousso:2020kmy}. In these prescriptions, de Sitter space is divided in three regions - the two static patches (in blue in Figure \ref{fig:dS}) and the exterior region between them (in white in Figure \ref{fig:dS}) - and extremal surfaces are computed in each region separately. The two prescriptions go as follows:
\vspace{0.2cm}

 $\bullet$ {\bf Monolayer proposal:} \textit{The entanglement entropy of a subregion $A$ of the two screens and its complement is $1/(4G\hbar)$ times the area of an extremal surface homologous\footnote{If there are multiple such extremal surfaces, we choose the one with the minimal area.} to $A$, and lying between the two sets of degrees
	of freedom; in this case between the two cosmological horizons.}
\vspace{0.2cm}

 $\bullet$ {\bf Bilayer proposal:} \textit{The entanglement entropy of a subregion $A$ of the two screens and its complement is $1/(4G\hbar)$ times the sum of the areas of the extremal surfaces homologous to $A$ in each of the three subregions of the bulk, that is, in the exterior and the two interior regions.}
 
 Here we argue briefly that the monolayer proposal appears inconsistent with the construction of the holographic description of de Sitter space presented above. For a detailed discussion, see \cite{Franken:2023pni}. The holographic description considered here starts from two complementary static patches encoded respectively on their cosmological horizons.  In particular, we already highlighted that whatever the state of two screen system $\mathcal{S}_{\rm L}\cup \mathcal{S}_{\rm R}$ is, it should at least encode the two static patches. The holographic entanglement entropy prescription should be consistent with this. As announced in the introduction of this paper, we suppose entanglement wedge reconstruction in the de Sitter background, with the associate usual properties of entanglement wedge nesting and homology constraint:
 \vspace{0.4cm}
 
 \textbf{Entanglement wedge reconstruction:} \textit{The bulk domain of dependence of the spacial slice whose boundary is the union of extremal surfaces in the exterior and the two interior regions (or only in the exterior region for the monolayer proposal) is reconstructible from the holographic subsystem $A$. This bulk domain of dependence is referred to as the entanglement wedge of $A$, noted $\mathcal{E}_A$. For $\mathcal{E}_A$ to be well defined, the covariant entropy prescription should define a set of surfaces, homologous to $A$ in their respective region, that are spacelike separated from any point of the holographic  screen and whose union is the boundary of a spacial region of the bulk.}
 
\vspace{0.4cm}
 By definition, the extremal surface always lies in the exterior region when considering the monolayer proposal. Hence, $\mathcal{E}_A$ lies entirely in the exterior region, and therefore never extends into the static patches. In particular, this implies that $\mathcal{E}_{\mathcal{S}_{\rm L}\cup \mathcal{S}_{\rm R}}$ does not cover the static patches, which is a contradiction. We will therefore discard the monolayer proposal in the rest of this work. We will give additional evidence against the monolayer proposal later in this work. In particular, the monolayer proposal fails to recover the prediction that the full de Sitter space is in a pure state \cite{Shaghoulian:2023odo, McNamara:2020uza, Marolf:2020xie, Almheiri:2019hni, Penington:2019kki}, and it also fails to apply to situations where the holographic screens are pushed on stretched horizon inside the static patch, contrary to the bilayer proposal.  In \cite{Franken:2023pni}, we carried out the definition of a covariant entropy prescription for both proposals, and showed additional inconsistent results in the monolayer case when considering non-trivial subsystems of the holographic screens. 
 
 One may argue that the absence of extremal surfaces in the static patch does not necessarily imply that entanglement wedges cannot cover them. However, in the absence of extremal surface in the static patches there is no preferred location for the entanglement wedge in these regions, such that the only other sensible possibility would be to state that the entanglement wedge of a subsystem $A$ of a screen always cover the associated static patch in addition to the region already defined. However this would also be contradictory as it leads to the conlusion that the entanglement wedge of a spatial subsystem of the horizon whose size goes to zero covers the full static patch. It would be interesting to explore the possibility that the monolayer proposal of \cite{Susskind:2021esx} gets supplemented by a consistent set of rules concerning the structure of the entanglement wedges, allowing them to extend in the interior regions. However, it is difficult to see how such a consistent set of rules could be formulated, since in all other well understood examples in the literature, the entanglement wedge structure depends on the location of the extremal surfaces, and the monolayer proposal does not involve any extremal surfaces in the interior regions. 

Following the HRT proposal of AdS/CFT, we would like to specify the bilayer proposal and define it in a covariant way. The prescription goes as follow \cite{Franken:2023pni}:\\

\noindent($i$) Consider an arbitrary Cauchy slice $\Sigma$ of dS$_{n+1}$.\\

\noindent($ii$) Define the screens $\mathcal{S}_{\rm L}$ and $\mathcal{S}_{\rm R}$ as the intersections of $\Sigma$ with $H_{\rm L}$ and $H_{\rm R}$, the cosmological horizons of the pode and the antipode, respectively. \\

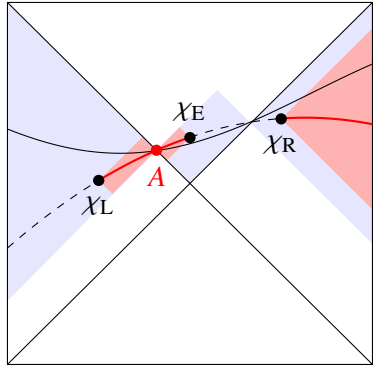
\begin{figure}[h!]
		\centering
		\begin{tikzpicture}[scale=0.8]
			\begin{scope}[transparency group]
					\path
					+(3,3)  coordinate (IItopright)
					+(-3,3) coordinate (IItopleft)
					+(3,-3) coordinate (IIbotright)
					+(-3,-3) coordinate(IIbotleft)
					
					;

					\fill[fill=blue!10] (-0.55,0.55) -- (0.45,1.55) -- (1,1) -- (0,0) -- cycle;
					
					\fill[fill=blue!10] (-3,3) -- (-0.55,0.55) -- (-3,-1.95) --  cycle;
					
					\fill[fill=blue!10] (1,1) -- (3,3) -- (3,-1) --  cycle;
					
					\fill[fill=red!30] (-1.5,0.05) -- (-0.775,0.775) -- (-0.55,0.55) --  (-1.275,-0.175) -- cycle;
					
					\fill [fill=red!30]  (-0.55,0.55) -- (-0.17,0.93) -- (0,0.76) --  (-0.38,0.38) -- cycle;

					\fill [fill=red!30] (1.5,1.07) -- (3,2.57) -- (3,-0.43) -- cycle;
					
					\draw (IItopleft) -- (IItopright) -- (IIbotright) -- (IIbotleft) --(IItopleft) -- cycle;
					
					\draw[domain=-3:-1.7, smooth, variable=\x] plot ({\x}, {sin(deg((\x/2-1)))+1.5});
					\draw[domain=-1.7:1.7, smooth, variable=\x, black] plot ({\x}, {sin(deg((\x/2-1)))+1.5});
					\draw[domain=1.7:3, smooth, variable=\x] plot ({\x}, {sin(deg((\x/2-1)))+1.5});

					\draw[domain=-3:-1.5, smooth, variable=\x, dashed] plot ({\x}, {1.09 - 0.09 *(-1.9 + \x)^2});
					\draw[domain=-1.5:0, smooth, variable=\x, red, thick] plot ({\x}, {1.09 - 0.09*(-1.9 + \x)^2});
					\draw[domain=0:1.5, smooth, variable=\x, dashed] plot ({\x}, {1.09 - 0.09*(-1.9 + \x)^2});
					\draw[domain=1.5:3, smooth, variable=\x, red, thick] plot ({\x}, {1.09 - 0.09*(-1.9 + \x)^2});

					\draw (IItopleft) -- (IIbotright)
					(IItopright) -- (IIbotleft) ;

					\node at (-0.55,0.55) [circle, fill, inner sep=1.5 pt, label = below:\color{red} $A$, red]{};
					
					\node at (-1.5,0.05) [circle, fill, inner sep=1.5 pt, label = below:$\chi_{\rm L}$]{};
					\node at (0,0.76) [circle, fill, inner sep=1.5 pt, label = above:$\chi_{\rm E}$]{};
					\node at (1.5,1.07) [circle, fill, inner sep=1.5 pt, label = below:$\chi_{\rm R}$]{};
					
			\end{scope}
		\end{tikzpicture}
		\caption{The Cauchy slice $\Sigma$ is the black curve and is cut in three parts by the horizons. $S_{\rm L}$ and $S_{\rm R}$ lie at the intersection between the black curve and the two diagonal lines. The red dot is a subregion of $S_{\rm L}\cup S_{\rm R}$ The region spanned by equivalent slices $\Sigma'$, equal to $D(\Sigma_{\rm L})\cup D(\Sigma_{\rm E})\cup D(\Sigma_{\rm R})$ is depicted in blue. Quantum extremal surfaces of each region are represented by black dots. All three belong to a Cauchy slice $\Sigma_i'$ depicted by a dashed line. The $\mathcal{C}_i$ are depicted by the red segments on $\Sigma'$. The entanglement wedge associated to the choice of extremal surfaces $\chi_i$ is the union of the three red regions.
			\label{fig:qucorr}}
\end{figure}

\noindent($iii$) Divide $\Sigma$ into 3 parts:
\begin{equation}
	\Sigma_i = \Sigma \cap P_i, \quad  \Sigma_{\rm E} = \Sigma \backslash \{\Sigma_{\rm L}\cup\Sigma_{\rm R}\},
\end{equation}
with $i=L,R$. Any Cauchy slice~$\Sigma'$ of dS$_{n+1}$  passing through $\mathcal{S}_{\rm L}$ and $\mathcal{S}_{\rm R}$ is equivalent to $\Sigma$, in the following sense. Decomposing in a similar way $\Sigma'=\Sigma'_{\rm L}\cup \Sigma'_{\rm E}\cup\Sigma'_{\rm R}$, the causal diamonds of $\Sigma'_{i}$ and $\Sigma_{i}$ in region $i$ are the same, for all $i\in\{\rm L, E, R\}$. See Figure \ref{fig:qucorr}.\\

\noindent($iv$) Consider any subregion $A$ of $\mathcal{S}_{\rm L}\cup \mathcal{S}_{\rm R}$. We first look for a surface~$\chi_{\rm E}$ of minimal extremal area that is homologous to $A_{\rm E}=A$ in the exterior region and is lying on a Cauchy slice $\Sigma'_{\rm E}$. We also look for a  surface $\chi_{\rm L}$ ($\chi_{\rm R}$) of minimal extremal area, homologous to $A_{\rm L}=A\cap \mathcal{S}_{\rm L}$ ($A_{\rm R}=A\cap \mathcal{S}_{\rm R}$) in the left (right) causal patch and lying on a Cauchy slice $\Sigma'_{\rm L}$ ($\Sigma'_{\rm R}$). For all $i\in\{\rm L,E,R\}$, recall that the codimension $2$ surface $\chi_i$ being homologous to the codimension $2$ surface $A_i$ means that there exists a codimension $1$ surface $\mathcal{C}_i$ such that $\partial \mathcal{C}_i =\chi_i \cup A_i$.\footnote{This implies in particular that $\chi_i$ is anchored on $A_i$, \ie $\partial \chi_i=\partial A_i$.\label{nb}} 
Among all possible choices of surfaces~$\mathcal{C}_i$, a particular one,  $\mathcal{C}'_i$, lies on~$\Sigma_i'$.  \\

\noindent($v$) At next-to-leading order in $G\hbar$, the Von Neumann entropy of the holographic dual subsystem on $A$, that is the entanglement entropy between $A$ and its complement in $\mathcal{S}_{\rm L}\cup \mathcal{S}_{\rm R}$, is
\begin{equation}
	\label{eq:entropy}
S(A)= \min \text{ext}_{\chi_i} \left[ \sum_{i}{\frac{{\rm Area}(\chi_i)}{4G\hbar}}\\ + S_{\rm semicl}\left(\bigcup_{j}{\cal {C}}'_{j} \right)\right] + \mathcal{O}\left((G\hbar)^1\right),
\end{equation}
 where $S_{\rm semicl}\left(\bigcup_{j}{\cal {C}}'_{j} \right)$ is the semiclassical entropy of bulk fields on the union of $\mathcal{C}'_i$ with $i=L,E,R$.\footnote{The necessity of joint extremization was already pointed out at the classical level in \cite{Shaghoulian:2021cef, Shaghoulian:2022fop}.} This choice is always minimal compared to the sum of the entropy of bulk fields on the three slices, by weak subadditivity. This is reminiscent of the island formula \cite{Almheiri:2020cfm}. \\
 
\noindent($vi$) The part in region $i$ of the ``entanglement wedge''  of the dual system living on $A$  is the causal diamond of~$\mathcal{C}'_i$. The causal diamond $D(\mathcal{C})$ of a spacelike region $\mathcal{C}$ is defined as the set of points $p$ of the bulk such that all timelike of lightlike curves that pass through $p$ must pass through $\mathcal{C}$. The entanglement wedge, called $\mathcal{E}(A)$, is therefore
\begin{equation}
\mathcal{E}(A) = D(\mathcal{C}'_{\rm L}) \cup D(\mathcal{C}'_{\rm E}) \cup D(\mathcal{C}'_{\rm R}).
\end{equation}

It was pointed out in \cite{Franken:2023pni} that for some examples of subsystem $A$, there may be no slice $\Sigma_i'$ in some region $i\in\{L,E,R\}$ such that there exist $\chi_i\in\Sigma_i$ an homologous extremal surface. However, whether extremal surfaces exist or not, non-extremal surfaces lying on the boundary of the causal diamond of $\Sigma_i$ can be seen as solutions of an extremization problem in an enlarged domain that has no boundary. In other words, we relax the definition of extremal surface $\delta \text{Area}(\chi_A)/\delta\chi_A=0$  and supplement the area functional with Lagrange multipliers that impose the surface to lie
inside a given region. Here we give an example of such implementation and refer to \cite{Franken:2023pni,Franken:2023jas} for more details and more complicated examples. We consider closed codimension $2$ surfaces in a $(n+1)$-dimensional de Sitter background and assume that the $SO(n)$ symmetry is not spontaneously broken when extremizing the
area functional. In other words, we look for $S^{n-1}$ spheres of extremal area in de Sitter space. The $(n+1)$-dimensional de Sitter admits the conformal metric
\begin{equation}
	ds^2 = \frac{1}{cos^2(\sigma)}\left(-d\sigma^2+d\theta^2+\sin^2(\theta)d\Omega_{n-1}^2\right),
	\label{eq:metric}
\end{equation}
where $\sigma\in(-\pi/2,\pi/2)$ and $\theta\in[0,\pi]$, which is explicitly $SO(n)$-symmetric. In these coordinates, every pair of coordinates $(\sigma,\theta)$ parameterize a $(n-1)$-sphere corresponding to a point in the Penrose diagram. The area of these spheres is given by
\begin{equation}
	\text{Area}(\sigma,\theta)=\omega_{n-1}\left(\frac{\sin\theta}{\cos\sigma}\right)^{n-1}
\end{equation}
where $\omega_{n-1}$ is the volume of the sphere of radius $1$. It is easy to show that the only extremal $S^{n-1}$, for which $\partial\text{Area}/\partial\sigma=\partial\text{Area}/\partial\theta=0$, is located at $(0,\pi/2)$. these coordinates correspond to the bifurcate horizon, the intersection between the cosmological horizons of the two observers. Hence, if we were to look for an extremal closed surface in a region $\mathcal{R}$ that does not contain the bifurcate horizon, there would be no solution to the extremization problem. Let us note $f_i(\sigma,\theta)\geq 0$ the set of inequations that define $\mathcal{R}$. We supplement the area functional with terms proportional to Lagrange multipliers:
\begin{equation}
	\label{eq:Lagrange}
 \widehat{\text{Area}}(\sigma,\theta,\nu_i,a_i) = \text{Area}(\sigma,\theta) + \sum_i  \nu_i \left(f_i(\sigma,\theta)-a_i^2\right).
\end{equation}
$\widehat{\text{Area}}$ is defined in a domain without boundary as $\sigma,\theta,\nu_i,a_i$ all span $\mathbb{R}$. Solving the problem of extremizing $\widehat{\text{Area}}$ over all these variables give the $S^{n-1}$ with extremal area with respect to $\mathcal{R}$, that is either a true extrema inside $\mathcal{R}$ (associated with $\nu_i=0$ for all $i$), or the sphere with the smallest or biggest area, or saddle points, inside $\mathcal{R}$ which will lie somewhere on the boundary $\partial\mathcal{R}$ (associated with $a_i=0$ for at least one $i$). If there are multiple such solutions, we are instructed to take the one with smallest area.

\section{Entanglement Wedge Reconstruction}
\label{sec:EW}

We now apply our covariant prescription to simple subsystems: The one-screen system and the two-screen system. In all instances, we always consider $SO(n)$-symmetric slices $\Sigma$ such that $\mathcal{S}_{\rm L}$ and $\mathcal{S}_{\rm R}$ are spherical screens. Let us denote $\sigma_{\rm L}$ and $\sigma_{\rm R}$ the conformal time of $\mathcal{S}_{\rm L}$ and $\mathcal{S}_{\rm R}$, respectively. One can re-express this in global time $\tau$ using 
\begin{equation}
 \cosh\tau=\frac{1}{\cos\sigma}.
\end{equation}
Since the screens lie on a Cauchy slice we always have $\tau_{\rm L}\tau_{\rm R}\geq 0$, or $\sigma_{\rm L}\sigma_{\rm R}\geq 0$. While we do expect each screen to encode its own static patch, it is not clear a priory if and how the exterior region is encoded in some of these subsystems. We end with a discussion on stretched horizons and the effects of integrating out degrees of freedom from the holographic screens.

\subsection{The Pair of Holographic Screens is Pure and Encodes Everything}
\label{sec:2screens}

The two-screen system is the simplest one to compute. We define $A=\mathcal{S}_{\rm L}\cup\mathcal{S}_{\rm R}$ and from our definitions of last section,
\begin{equation}
 A_{\rm L} = \mathcal{S}_{\rm L},\quad A_{\rm E}= \mathcal{S}_{\rm L}\cup\mathcal{S}_{\rm R}, \quad  A_{\rm R} = \mathcal{S}_{\rm R}.
\end{equation}
All these surfaces are closed. Hence, in each region $i=L,E,R$ we are looking for an extremal surface without boundary lying on a Cauchy slice $\Sigma'_i$ as defined in the third step of our procedure. The empty surface defined by the empty set of points is the subset of any Cauchy slice $\Sigma'_i$ and has minimal extremal area by definition. It satisfies the homology constraint in every region,
\begin{equation}
	A_i \cup \varnothing = \partial \mathcal{C}'_i,
\end{equation}
where $\mathcal{C}'_i$ is any of the Cauchy slices $\Sigma'_i$ as defined in the third step of our procedure. This is verified by definition of $\Sigma'_i$. Hence, the entropy of the two-screen system, consisting of some spatial slice of the full dual theory, is given by
 \begin{equation}
	S(\mathcal{S}_{\rm L}\cup\mathcal{S}_{\rm R})= 0+  \mathcal{O}\left((G\hbar)^1\right),
\end{equation}
where we used $\text{Area}(\varnothing)=0$, $\Sigma'=\Sigma'_{\rm L}\cup\Sigma'_{\rm E}\cup\Sigma'_{\rm R}$, and $S_{\rm semicl}\left(\Sigma' \right)=0$ since $\Sigma'$ is a complete Cauchy slice. In other words, the full holographic system has zero entropy at next-to-leading order. Following the covariant prescription, the part of the entanglement wedge in each region $i=L,E,R$ is given by the causal diamond of $\mathcal{C}_i'=\Sigma'_i$, such that
\begin{equation}
	\mathcal{E}(\mathcal{S}_{\rm L}\cup\mathcal{S}_{\rm R}) = D(\Sigma_{\rm L}) \cup D(\Sigma_{\rm E}) \cup D(\Sigma_{\rm R}),
\end{equation} 
where we used $D(\Sigma'_i)=D(\Sigma_i)$. This region is spanned by equivalent complete slices $\Sigma'$. In other words, $\mathcal{S}_{\rm L}\cup\mathcal{S}_{\rm R}$ encodes all complete Cauchy slices passing through $\mathcal{S}_{\rm L}\cup\mathcal{S}_{\rm R}$. The fact that the entanglement wedge of the two holographic screens covers the exterior region in addition to their static patches leads to an extension of the statement of static patch holography: 

\paragraph{Doubled static patch holography}
\emph{The full de Sitter spacetime can be encoded
	holographically in terms of a theory on the two cosmological horizons.\footnote{Or on stretched horizons separated from the cosmological horizon by less than a Planck length, see Section \ref{sec:stretched}.} The states of the interior regions are encoded on the screens at the corresponding horizons, as in static patch holography. Additionally, the exterior region between the cosmological horizons emerges from the entanglement between the screens at the horizons and can also be reconstructed from the holographic degrees of freedom.}\\

This story is reminiscent of the double-sided black hole in AdS as already highlighted  in Section \ref{sec:bridge}. Both screens have Hilbert space $\mathcal{H}_{L(R)}$ and encode their respective static patch $P_{L(R)}$. Once we consider the full theory of Hilbert space $\mathcal{H}_{\rm L}\otimes\mathcal{H}_{\rm R}$, the entanglement between the two sets of degrees of freedom can be seen from the bulk point of view as the spatial connectivity between the two patches $P_{L(R)}$. In other words, the entanglement wedge of the union of the two screens is bigger than the union of their independent entanglement wedges, and it covers the full exterior region that bridges the two static patches. This enrichment of the bulk geometry emerges directly from entanglement in the dual theory. 

\begin{figure}[!h]
\begin{subfigure}[t]{0.48\linewidth}
	\centering
	\begin{tikzpicture}[scale=0.5]
		\begin{scope}[transparency group]
			\begin{scope}[blend mode=multiply]
				\path
				+(3.5,3)  coordinate (IItopright)
				+(-3.5,3) coordinate (IItopleft)
				+(3.5,-3) coordinate (IIbotright)
				+(-3.5,-3) coordinate(IIbotleft)
				
				;
				\draw (IItopright) --
				(IIbotright); 
				\draw (IIbotleft) --
				(IItopleft);

				
				
				\fill[fill=blue!20] (-3.5,3) -- (-3.5,-3) -- (-0.5,0);
				
				\fill[fill=blue!20] (3.5,3) -- (3.5,-3) -- (0.5,0);
				
				\draw[line width=0.5mm] (IItopleft)  -- node[midway, above, sloped] {Left screen} (-0.5,0) -- (IIbotleft);
				\draw[line width=0.5mm] (IItopright) -- node[midway, above, sloped] {Right screen}  (0.5,0)--(IIbotright) ;
				
				\node at (0,0) [label = center:$\otimes$]{};
			\end{scope}
		\end{scope}
	\end{tikzpicture}
\caption{}
\end{subfigure}
\begin{subfigure}[t]{0.48\linewidth}
	\centering
	\begin{tikzpicture}[scale=0.5]
		\begin{scope}[transparency group]
			\begin{scope}[blend mode=multiply]
				\path
				+(3,3)  coordinate (IItopright)
				+(-3,3) coordinate (IItopleft)
				+(3,-3) coordinate (IIbotright)
				+(-3,-3) coordinate(IIbotleft)
				
				;
				\draw (IItopleft) --
				(IItopright) --
				(IIbotright) -- 
				(IIbotleft) --
				(IItopleft) -- cycle;

				\fill[fill=red!20] (-3,3) -- (3,3) -- (0,0);
				
				\fill[fill=red!20] (-3,-3) -- (3,-3) -- (0,0);
				
				\fill[fill=blue!20] (-3,3) -- (-3,-3) -- (0,0);
				
				\fill[fill=blue!20] (3,3) -- (3,-3) -- (0,0);
				
				\draw[line width=0.5mm] (IItopleft) -- (IIbotright)
				(IItopright) -- (IIbotleft) ;
			\end{scope}
		\end{scope}
	\end{tikzpicture}
\caption{}
\end{subfigure}
\caption{Extended static patch holography: (a) We consider initially two independant static patches, independently encoded on two holographic screens of Hilbert space $\mathcal{H}_{\rm L}$ and $\mathcal{H}_{\rm R}$. (b) Once we take into account the entanglement between these two theories, described by a Bunch-Davies state on a Hilbert space $\mathcal{H}_{\rm L}\otimes\mathcal{H}_{\rm R}$, the full bulk dual theory appears to be the complete de Sitter space. This picture is analogous to the AdS black hole presented in Figure \ref{fig:ER=EPR}.\label{fig:bridge}}
\end{figure}
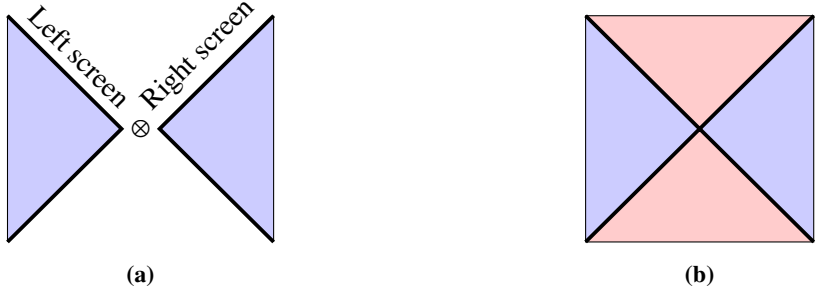

Similarly to the double-sided AdS black hole, the state of both the dual quantum theory and the state of the de Sitter space can be described by a thermofield-double state called the Bunch-Davies state \cite{Goheer:2002vf}, constructed as follows. We consider a geodesic in de Sitter space, which is arbitrary and equivalent to any other one thanks to the isometry group of de Sitter space, and place the worldline of an observer along this geodesic. This worldline defines a static patch. Among the Killing vectors of de Sitter space, there is one that is timelike, future-directed, and defined everywhere in the static patch. This killing vector maps the worldline to itself, by shifting the proper time on the worldline. The associated conserved charge $H$ is the Hamiltonian of the static patch.\footnote{In \cite{Witten:2023xze}, this Hamiltonian is supplemented by the Hamiltonian of the observer itself.} Let us write $\ket{\Psi_i}$ and $E_i$ are the eigenvectors and eigenvalues of $H$. We know that the static patch is in a thermal state of inverse temperature $\beta$:
\begin{equation}
	\label{eq:th}
 \rho = \frac{1}{Z}e^{-\beta H}.
\end{equation}
We consider two copies of such static patch, with Hilbert spaces $\mathcal{H}_{\rm L}$ and $\mathcal{H}_{\rm R}$ and associated eigenstates $\ket{\Psi_i}_{\rm L}$ and $\ket{\Psi_i}_{\rm R}$. Each of these theories independently describes the static patch, they correspond to the left picture of Figure \ref{fig:bridge} where the two static patches are completely independent. 

We now construct the thermofield-double state whose Hilbert space is $\mathcal{H}_{\rm L}\otimes \mathcal{H}_{\rm R}$ and define the Hamiltonian
\begin{equation}
	H_{\rm BD}= H_{\rm L} - H_{\rm R},
\end{equation}
with $H_{\rm L}$ and $H_{\rm R}$ two copies of the original Hamiltonian. Considering physical states of $\mathcal{H}_{\rm L}\otimes \mathcal{H}_{\rm R}$ with vanishing eigenvalues ensures that both static patches have the same time. This can be seen as one of the static patch playing the role of a clock for the other \cite{Narovlansky:2023lfz}. With this constraint in mind, one constructs the thermofield-double state,
\begin{equation}
 \ket{\Psi_{\rm BD}} = \frac{1}{\sqrt{Z}}\sum e^{-\frac{1}{2}\beta E_i}\ket{\Psi_i}_{\rm L}\otimes \ket{\Psi_i}_{\rm R},
\end{equation}
which satisfies the constraint $H_{\rm BD} \ket{\Psi_{\rm BD}}=0$. Tracing out the degrees of freedom from $\mathcal{H}_{\rm L}$ ($\mathcal{H}_{\rm R}$), one recovers the thermal density matrix \eqref{eq:th} associated with $P_{\rm L}$ ($P_{\rm R}$). The Bunch-Davies state corresponds to the full de Sitter space as in the right panel of Figure \ref{fig:bridge}. In this picture, the two static patches in blue are still encoded in the Hilbert spaces $\mathcal{H}_{\rm L}$ and $\mathcal{H}_{\rm R}$, while the exterior region in red is encoded in $\ket{\Psi}_{\rm BD}$, emerging from the entanglement between $\mathcal{H}_{\rm L}$ and $\mathcal{H}_{\rm R}$.  It was argued in \cite{Chandrasekaran:2022cip} that in a quantum description of de Sitter space, the static patches must be maximally entangled which is associated to $\beta\rightarrow 0$.  This might seem counter-intuitive as we know that the de Sitter background can be associated with a finite temperature $1/\beta_{dS}=2\pi$ where the de Sitter radius has been fixed to $1$. But it was argued in \cite{Lin:2022nss} that it is possible to define a temperature-like quantity, named tomperature, which stays finite as the global temperature goes to infinity.

The Bunch-Davies state is pure, which is consistent with the discussion of Section \ref{sec:observer} and is consistent with the idea that the state of the whole universe should be unique and pure. The entropy of the pair of holographic screens is therefore expected to be pure not only at next-to-leading order but to all orders. In case one would doubt that the exterior region is encoded on the screens, one could argue that when the screens coincide at the bifurcate horizons there is no exterior region such that the two-screen system must be in a pure state. Indeed, there exist Cauchy slices $\Sigma'$ such that $\Sigma'_{\rm L }$ and $\Sigma'_{\rm R}$ join at the bifurcate horizon, such that $\Sigma'_{\rm E}=\varnothing$. Moreover, the evolution operator along the screens should be linear (possibly unitary), which cannot transform a pure state into a mixed state. Hence, whether we believe or not that the exterior region is also encoded on the screens, the entropy of the two-screen system must vanish to all orders:
\begin{equation}
	S(\mathcal{S}_{\rm L}\cup\mathcal{S}_{\rm R})= 0 +  \mathcal{O}\left((G\hbar)^n\right),
\end{equation}
for any $n\geq0$.
The discussion above is another argument against the monolayer proposal. In the monolayer proposal, extremal surfaces do not extend in the static patch such that we do not expect the entanglement wedge of the screens to extend in them either (see Section \ref{sec:ent}). Beyond the fact that this is inconsistent with static patch holography, one may note that this implies that non vanishing $\mathcal{O}((G\hbar)^0)$-terms would appear in $S(\mathcal{S}_{\rm L}\cup\mathcal{S}_{\rm R})$ in the monolayer proposal. In particular, the correction term in equation \eqref{eq:entropy} would be modified to $S_{\text{semicl}}(\mathcal{C}'_{\rm E})$ which does not vanish in general.

\subsection{Single-Screen System: A Phase Transition of $\mathcal{E}(\mathcal{S}_{\rm L})$}

In this section we consider the simplest non-trivial subsystem of the full holographic system, that is $\mathcal{S}_{\rm L}$ or $\mathcal{S}_{\rm R}$. Here we discuss $\mathcal{S}_{\rm L}$. Since the two-screen system has zero entropy,
\begin{equation}
 S(A)=S(\bar{A}),
\end{equation}
 where $\bar{A}$ is the complementary subsystem with respect to $\mathcal{S}_{\rm L}\cup \mathcal{S}_{\rm R}$. Hence our computations for any susbsystem directly transfer to its complement. In particular, we can also expect $\chi_{\rm E}(A) =  \chi_{\rm E}(\bar{A}) $.
 
In the left static patch, $A_{\rm L}=\mathcal{S}_{\rm L}$ and $\varnothing$ is the minimal extremal surface with $\mathcal{C}'_{\rm L}=\Sigma'_{\rm L}$ for the same reason as in the previous section. The part of the entanglement wedge lying in the left static patch therefore covers $\Sigma'_{\rm L}$, which is exactly what we expect from static patch holography. On the other hand, $A_{\rm R}=\varnothing$ which is homologous to $\varnothing$ with $\mathcal{C}'_{\rm R}=\varnothing$. The empty surface $\varnothing$ being of extremal and vanishing area, it is the minimal extremal homologous surface in the right static patch. The causal diamond of $\varnothing$ being $\varnothing$, the entanglement wedge of $\mathcal{S}_{\rm L}$ does not extend in the right static patch. 

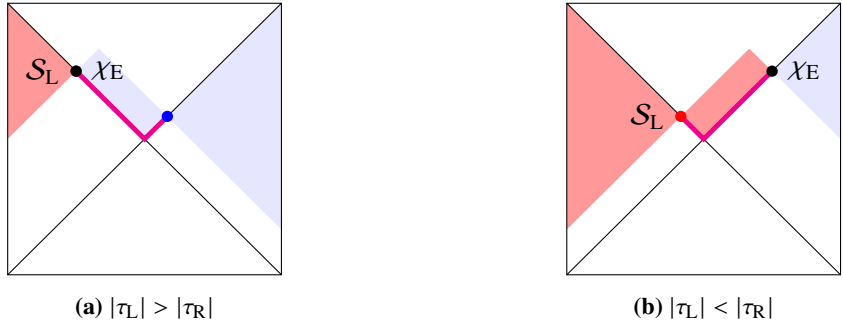
\begin{figure}[h!]
	\begin{subfigure}[t]{0.48\linewidth}
		\centering
		\begin{tikzpicture}[scale=0.6]
			\begin{scope}[transparency group]
					\path
					+(3,3)  coordinate (IItopright)
					+(-3,3) coordinate (IItopleft)
					+(3,-3) coordinate (IIbotright)
					+(-3,-3) coordinate(IIbotleft)
					
					;		
					\fill[fill=blue!10] (-3/2,3/2) -- (-1,2) -- (1/2,1/2) -- (0,0) -- cycle;
					
					\fill[fill=red!40] (-3,3) -- (-3/2,3/2) -- (-3,0) --  cycle;
					
					\fill[fill=blue!10] (1/2,1/2) -- (3,3) -- (3,-2) --  cycle;

					\draw (IItopleft) --
					(IItopright) --
					(IIbotright) -- 
					(IIbotleft) --
					(IItopleft) -- cycle;

					\draw (IItopleft) -- (IIbotright)
					(IItopright) -- (IIbotleft) ;
						
					\draw[line width=0.65mm, color=magenta] (-3/2,3/2) -- (0,0) -- (1/2,1/2);

					\node at (-3/2,3/2) [circle,fill,inner sep=1.5pt, red, label = left:$\mathcal{S}_{\rm L}$]{};
					\node at (-3/2,3/2) [circle,fill,inner sep=1.5pt, black, label = right:$\chi_{\rm E}$]{};
					\node at (1/2,1/2) [circle,fill,inner sep=1.5pt, blue]{};
			\end{scope}
		\end{tikzpicture}
		\caption{$ |\tau_{\rm L}|>|\tau_{\rm R}|$}
	\end{subfigure}
\begin{subfigure}[t]{0.48\linewidth}
	\centering
\begin{tikzpicture}[scale=0.6]
	\begin{scope}[transparency group]
	\path
	+(3,3)  coordinate (IItopright)
	+(-3,3) coordinate (IItopleft)
	+(3,-3) coordinate (IIbotright)
	+(-3,-3) coordinate(IIbotleft)
	
	;
	
	\fill[fill=red!40] (-1/2,1/2) -- (1,2) -- (3/2,3/2) -- (0,0) -- cycle;
	
	\fill[fill=red!40] (-3,3) -- (-1/2,1/2) -- (-3,-2) --  cycle;
	
	\fill[fill=blue!10] (3/2,3/2) -- (3,3) -- (3,0) -- cycle;
	
	\draw (IItopleft) --
	(IItopright) --
	(IIbotright) -- 
	(IIbotleft) --
	(IItopleft) -- cycle;
	
	\draw (IItopleft) -- (IIbotright)
	(IItopright) -- (IIbotleft) ;
	
	
	
	\draw[line width=0.65mm, color=magenta] (3/2,3/2) -- (0,0) -- (-1/2,1/2);
	
	\node at (-1/2,1/2) [circle,fill,inner sep=1.5pt, red, label = left:$\mathcal{S}_{\rm L}$]{};
	\node at (3/2,3/2) [circle,fill,inner sep=1.5pt, black, label = right:$\chi_{\rm E}$]{};
\end{scope}
\end{tikzpicture}
\caption{$ |\tau_{\rm L}|<|\tau_{\rm R}|$}
\end{subfigure}
\caption{The sceen $\mathcal{S}_{\rm L}$ of which we compute the entanglement entropy is depicted by a red point while its entanglement wedge $\mathcal{E}(\mathcal{S}_{\rm L})$ is the red region. The complementary screen $\mathcal{S}_{\rm R}$ is the blue dot. The extremal surfaces in the two interior regions are $\varnothing$ while the non-trivial extremal surface in the exterior, $\chi_{\rm E}$, is depicted by a black dot. The light blue region indicates the region spanned by all $\Sigma'$ and which is not contained in $\mathcal{E}(\mathcal{S}_{\rm L})$. This blue region also corresponds to $\mathcal{E}(\mathcal{S}_{\rm R})$. The thick magenta line corresponds to the union of all candidate extremal surfaces with minimal area $\text{Area}(\mathcal{S}_{\rm L})$.
	\label{fig:phase}}
\end{figure}

Since the two interior regions do not contribute to the entropy, the entropy of $\mathcal{S}_{\rm L}$ entirely comes from the exterior region, as one would expect from the ER=EPR correspondence. Here, $A_{\rm E}=\mathcal{S}_{\rm L}$ has no boundary but the empty surface is not homologous since there is not any $\Sigma'_{\rm E}$ that contains a spatial system only bounded by $\mathcal{S}_{\rm L}$. Hence, we look for a non-trivial $SO(n)$-symmetric surface in the exterior that has minimal extremal area.\footnote{We suppose that the $SO(n)$ symmetry is not spontaneously broken during the extremization of the area functional.} While the bifurcate horizon is the only extremal surface in the exterior region, every $SO(n)$-symmetric slice of the horizon has the same area as the bifurcate horizon, see Figure \ref{fig:phase}. In \cite{Franken:2023pni}, we showed that all such surfaces in the exterior region are solutions of the extremization problem \eqref{eq:Lagrange} and should be considered as candidates in our holographic entropy prescription. At leading order, these surfaces give degenerate results:
 \begin{equation}
 	S(\mathcal{S}_{\rm L}) = \frac{\text{Area}(\mathcal{S}_{\rm L})}{4G\hbar} + \mathcal{O}((G\hbar)^0),
 \end{equation}
Hence we recover the result of \cite{Susskind:2021omt, Shaghoulian:2021cef,Shaghoulian:2022fop} that the entanglement entropy of one screen is equal to the Bekenstein-Hawking formula at leading order.

However, these degenerate extremal surfaces lead to completely different entanglement wedges. In particular, extremal surfaces lying on $H_{\rm L}$ lead to an entanglement wedge that does not extend in the exterior but only in the interior and along the horizon. On the other hand, extremal surfaces lying on $H_{\rm R}$ lead to an entanglement wedge that extends in the exterior region. The most extreme case being when the extremal surface lies on $\mathcal{S}_{\rm R}$. In this case, $\mathcal{C}'_{\rm E}=\Sigma'_{\rm E}$ and the entanglement wedge connects the two holographic screens. 

This degeneracy may be lifted in two ways. The first one is to consider classical perturbations of the metric and was studied in \cite{Franken:2023jas}. The second one, which we consider in this work, is to take into account the next-to-leading order contribution to the entropy. The entropy of $\Sigma'_{\rm L}$ decreases as $\mathcal{S}_{\rm L}$ evolves in (absolute value of) time along $H_{\rm L}$. Both screens $\mathcal{S}_{\rm L}$ and $\mathcal{S}_{\rm R}$ are always favored in the extremization of the generalized entropy \eqref{eq:entropy}, as they are the $SO(n)$-symmetric slices of the horizons at the highest global time $|\tau|$ (and hence the minimal semiclassical correction) in the region spanned by $\Sigma'_i$.\footnote{As already noted, $SO(n)$-symmetric surfaces are associated with a unique value of $\tau$ and $\theta$.} The procedure then simply instructs us to choose the one with the smallest semiclassical correction, \textit{i.e.} with the highest value of $|\tau|$. Hence,
\begin{equation}
\chi_{\rm E} = \left\{
\begin{array}{ll}
	\mathcal{S}_{\rm L} & \mbox{if }| \tau_{\rm L}|>|\tau_{\rm R}| \\
	\mathcal{S}_{\rm R} & \mbox{if } |\tau_{\rm L}|<|\tau_{\rm R}| 
\end{array}
\right., \qquad \mathcal{E}(\mathcal{S}_{\rm L}) = \left\{
\begin{array}{ll}
	D(\Sigma_{\rm L}) & \mbox{if } |\tau_{\rm L}|>|\tau_{\rm R}| \\
	D(\Sigma_{\rm L})\cup D(\Sigma_{\rm E}) & \mbox{if} |\tau_{\rm L}|<|\tau_{\rm R}| 
\end{array}
\right.,
\end{equation}
with a phase transition at $\tau_{\rm L}=\tau_{\rm R}$. In this case, higher order corrections are needed to lift the degeneracy. As for the entropy, we get
\begin{equation}
S(\mathcal{S}_{\rm L}) = \frac{\text{Area}(\mathcal{S}_{\rm L})}{4G\hbar} +  \left\{
\begin{array}{ll}
	S_{\text{semicl}}(\Sigma'_{\rm L}) & \mbox{if } |\tau_{\rm L}|>|\tau_{\rm R}| \\
	S_{\text{semicl}}(\Sigma'_{\rm R}) & \mbox{if } |\tau_{\rm L}|<|\tau_{\rm R}| 
\end{array}
\right.,
\end{equation}
where we used $S_{\text{semicl}}(\Sigma'_{\rm R})=S_{\text{semicl}}(\Sigma'_{\rm L}\cup\Sigma'_{\rm E})$. The entanglement wedges and extremal surfaces are depicted in Figure \ref{fig:phase}.

We thus showed that, while the full holographic system encodes the whole de Sitter space, there is always one of the two screens that contains enough information to encode both the interior of his static patch and the exterior region. The complementary screen then only encodes its own interior. The screen that encodes the exterior is the one with the smallest entanglement entropy. Another way to see this is to consider the Quantum Bousso Bound (QBB) of Strominger and Thompson \cite{Strominger:2003br}. While the classical Bousso Bound relates the number of degrees of freedom of a region to an area, the Strominger-Thompson QBB which was shown to be valid in JT gravity \cite{Franken:2023ugu} associates these degrees of freedom to a quantum area
\begin{equation}
 A_{\text{qu}}(\chi)=\text{Area}(\chi) + 4G\hbar S_{\text{semicl}}(\Sigma_\chi), 
\end{equation}
where $\Sigma_{\chi}$ is a spacelike slice bounded by $\chi$. From this point of view, we can see the screen with the smallest entanglement entropy as the screen with the biggest quantum area and therefore containing the largest number of degrees of freedom, thus dominating the competition for the encoding of the exterior region. 

At $\tau_{\rm L}=\tau_{\rm R}$, there is a phase transition where the screen containing the greatest number of degrees of freedom switch from one to the other. This results in an exchange of dominance between the two competing quantum extremal surfaces. This phase transition reflects into the entanglement wedge picture as the encoding of the exterior region transfers from one screen to the other. In \cite{Engelhardt:2023xer}, this type of transition in AdS/CFT was identified with the transfer of an emergent type III$_1$ subalgebra whose reconstruction complexity is exponential in Newton's constant, called complex factor, from one screen subsystem algebra to the other. It would be very interesting to understand how the analysis of \cite{Engelhardt:2023xer} would apply in this context.

\subsection{Pushing the Screens Inside the Static Patches}
\label{sec:stretched}

Lastly, we would like to comment on the effect of pushing the holographic screens inside the the static patch. Let us work in static patch coordinates,
\begin{equation}
	ds^2 = -(1-r^2)dr^2 + \frac{dr^2}{1-r^2}+ r^2 d\Omega_{n-1}^2,
\end{equation}
where we have set the de Sitter radius to $1$. We consider the stretched horizon as the timelike hypersurface defined by $r=1-\epsilon$. In general stretched horizons are defined such that $\epsilon\sim l_{\rm Pl}$ with $l_{\rm Pl}$ the Planck length. But we would like to relax this definition for the moment to consider hypersurfaces lying further from the cosmological horizon. A screen placed on the stretched horizon therefore evolves along the hyperbola $r=1-\epsilon$ and its area remains constant, decreased by a factor $(1-\epsilon)^{n-1}$ compared to the parent screen on the cosmological horizon. The temperature measured in the static patch is given by
\begin{equation}
 T_{\rm dS} = \frac{1}{2\pi\sqrt{1-r^2}},
\end{equation}
in units of the de Sitter radius. The stretched horizon therefore acts as a cutoff, with the de Sitter temperature varying from $1/(2\pi)$ along the observer trajectory, to a temperature scaling as $\epsilon^{-\frac{1}{2}}$.  

In Section \ref{sec:dSholo}, we argued that putting the holographic screen on the cosmological horizon maximizes the number of holographic degrees of freedom. But one could decide to stay away from the mathematical horizon and place the holographic screen on the hypersurface $r=1-\epsilon$ in the static patch, as in \cite{Susskind:2021omt}. This is usually the prescription followed when using static patch holography as it plays the role of a cutoff. However, in the context of entanglement wedge reconstruction, this leads to radical issues. For example, applying blindly our covariant prescription would lead to the same conclusion that the region between the two stretched horizon are reconstructible from the holographic degrees of freedom located on them. However, at a rate of $1$ degree of freedom per Planck area, the number of holographic degrees of freedom decreases to zero as the stretched horizons is pushed towards the podes, \textit{i.e.} as $\epsilon\rightarrow 1$. This seems paradoxical. 

Here, we would like to argue that while static patch holography stays consistent when the holographic screen is pushed inside the static patch, the bilayer proposal must be reconsidered. Pushing the screens in the static patches effectively integrates out holographic degrees of freedom, as the area of the screen decreases by a factor $(1-\epsilon)^{n-1}$. In particular, this change of location can be seen as replacing the holographic theory by a cutoff version of the initial one, where the degrees of freedom encoding the two strip regions bounded by the stretched horizons and their associated cosmological horizon have been integrated out. 

Moreover, we expect that with this loss of degrees of freedom, the entanglement between the two holographic theories is no longer sufficient to reconstruct the exterior region between the two cosmological horizons. In other words, the entanglement wedges of subsystems of this cutoff theory can no longer extend in the region between the two holographic screens. This can be seen as follows. As the strips regions are effectively removed from the holographic duality, the screens can no longer be deformed in the exterior region. Indeed, extremal surfaces should be seen as continuous deformations of the screens subsystem into the holographic bulk region. But removing the two strips completely disconnect the two static patches from the exterior, such that extremal surfaces and entanglement wedges cannot extend in this region. Hence, the entanglement entropy of one screen should only receive a semiclassical contributions from the fields in its interior.

The two cutoff theories can be unitarily evolved from the pair of screens at the symmetric slice $\tau=0$, as their area is constant in time. However, the full holographic theory is not expected to be a pure state as the entropy of the two-screen system is no longer expected to vanish to all orders in $\hbar$, since there cannot be contributions from the bulk region  between the screens. The discussion above should be restricted to semiclassical gravity, such that variations of the location of the holographic screens at the Planck scale or below should not be taken seriously. In particular, a screen located on the stretched horizon with $\epsilon$ scaling as $\mathcal{O}(\hbar^{a})$ with $a>1$ would not get its area decreased by more than one Planck area. In other words, a stretched horizon sufficiently close to the cosmological horizon will not have less holographic degrees of freedom. For this reason, we restrict our conclusions concerning the loss of the degrees of freedom to screens that are separated from the cosmological horizon by more than a Planck length, further inside than the stretched horizon with $\epsilon\sim l_{\rm Pl}$.
 
 \section*{Acknowledgements}
I am grateful to Hervé Partouche, Nicolaos Toumbas, and in particular to Lior Benizri and François Rondeau for helpful discussions and feedback on the manuscript. This work is partially supported by the Cyprus Research and Innovation Foundation grant EXCELLENCE/0421/0362.
 

\bibliographystyle{jhep}

\bibliography{skeleton}

\end{document}